\documentclass[utf8]{frontiersSCNS} 

\usepackage{url,hyperref,lineno,microtype,subcaption}
\usepackage[onehalfspacing]{setspace}

\DeclareRobustCommand{\ion}[2]{\textup{#1\,\textsc{\lowercase{#2}}}}
\newcommand*\aap{A\&A}

\newcommand*\aj{AJ}

\newcommand*\apj{ApJ}
\newcommand*\apjl{ApJ}

\newcommand*\araa{ARA\&A}

\newcommand*\solphys{Sol.~Phys.}

\newcommand*\ssr{Space~Sci.~Rev.}

\newcommand*\arcsec{\ensuremath{^{\prime\prime}}}

\def\keyFont{\fontsize{8}{11}\helveticabold }
\def\firstAuthorLast{da Silva Santos {et~al.}} 
\def\Authors{J. M. da Silva Santos\,$^{1,*}$, S. M. White\,$^{2}$, K. Reardon\,$^{1}$, G. Cauzzi\,$^{1}$, S. Gunár\,$^{3}$,\\ P. Heinzel\,$^{3}$, and J. Leenaarts\,$^{4}$}
\def\Address{$^{1}$National Solar Observatory, Boulder, CO, USA\\
$^{2}$Space Vehicles Directorate, Air Force Research Laboratory, Albuquerque, NM, USA\\$^{3}$Astronomical Institute of the Czech Academy of Sciences, Ondřejov, Czech Republic\\$^{4}$Institute for Solar Physics, Department of Astronomy, Stockholm University, Stockholm, Sweden}
\def\corrAuthor{3665 Discovery Drive, Boulder 80303, CO, USA}

\def\corrEmail{jdasilvasantos@nso.edu}

\begin{document}
\onecolumn
\firstpage{1}

\title[Thermal structure of AR filaments]{Subarcsecond imaging of a solar active region filament with ALMA and IRIS} 

\author[\firstAuthorLast ]{\Authors} 
\address{} 
\correspondance{} 

\extraAuth{}

\maketitle

\begin{abstract}

Quiescent filaments appear as absorption features on the solar disk when observed in chromospheric lines and at continuum wavelengths in the millimeter (mm) range. 
Active region (AR) filaments are their small-scale, low-altitude analogues, but they could not be resolved in previous mm observations. This spectral diagnostic can provide insight into the details of the formation and physical properties of their fine threads, which are still not fully understood.
Here, we shed light on the thermal structure of an AR filament using high-resolution brightness temperature ($T_{\rm b}$) maps taken with ALMA Band\,6 complemented by simultaneous IRIS near-UV spectra, Hinode/SOT photospheric magnetograms, and SDO/AIA extreme-UV images. 
Some of the dark threads visible in the AIA 304\,\AA~passband and in the core of \ion{Mg}{II} resonance lines have dark ($T_{\rm b}$\,$<$\,$5000$\,K) counterparts in the 1.25\,mm maps, but their visibility significantly varies across the filament spine and in time. These opacity changes are possibly related to variations in temperature and electron density in filament fine structures. The coolest $T_{\rm b}$ values ($<$\,$5000$\,K) coincide with regions of low integrated intensity in the \ion{Mg}{II} h and k lines. ALMA Band\,3 maps taken after the Band\,6 ones do not clearly show the filament structure, contrary to the expectation that the contrast should increase at longer wavelengths based on previous observations of quiescent filaments. The ALMA maps are not consistent with isothermal conditions, but the temporal evolution of the filament may partly account for this.

\tiny
 \keyFont{ \section{Keywords:} Sun, radio, ultraviolet, chromosphere, active regions, filaments, prominences} 
\end{abstract}

\newpage
\section{Introduction}

Dark filaments on the solar disk and bright prominences at the limb are homologous structures filled with cool ($\lesssim$\,$10^{4}$\,K), dense plasma suspended by the Lorentz force against gravity above polarity-inversion lines in the photosphere. They are generally separated into two categories: the larger ($\sim$10\,$-$\,100\,Mm), long-lived, high-altitude ($>$\,10\,Mm) quiescent filaments and the smaller ($\sim$10\,Mm), more dynamic, lower-altitude ($<$\,10\,Mm) active region (AR) filaments. 
The magnetic field in the former is essentially horizontal with field strengths of $\sim$3-80\,G \citep[][]{1989ASSL..150...77L,2005ApJ...622.1265C,2014A&A...566A..46O}, but AR filaments may show flux rope structure and field strengths that can be an order of magnitude higher than in the quiescent ones \citep{2009A&A...501.1113K,2012A&A...539A.131K,2010ApJ...714..343G,2012ApJ...749..138X,2019A&A...623A.178D}. Estimates of (core) kinetic temperatures in prominences lie within the range $\sim$\,5\,000\,$-$\,9\,000\,K for microturbulence values within $\sim$\,3\,$-$\,16\,$\rm km\,s^{-1}$ \citep[e.g.,][]{2003SoPh..217..133S,2018A&A...618A..88J,2019ApJ...886..134R}.
While the general properties of filaments are well understood, there are open questions pertaining to their fine structure and magnetic configuration \citep[][and references therein]{2020RAA....20..166C}. Furthermore, they may become unstable and erupt, leading to coronal mass ejections, hence their relevance in the context of solar activity and space weather \citep[see reviews by][]{2014LRSP...11....1P,2018LRSP...15....7G}.

Observational studies have mainly used spectral lines such as \ion{Ly}{$\alpha$}, \ion{H}{$\alpha$}, \ion{Ca}{II}\,H, 8542\,\AA, \ion{He}{I}\,D$_{3}$, 10830\,\AA, and \ion{He}{II}\,304\,\AA~to investigate prominence/filament thermodynamics and to infer the magnetic field topology \citep[e.g.,][and references therein]{2010SSRv..151..243L}.
These structures have also been routinely observed in the last decade by the Interface Region Imaging Spectrograph \citep[IRIS,][]{2014SoPh..289.2733D} in the \ion{Mg}{II} h and k resonance lines, which have been used to investigate dynamics in prominences \citep[e.g.][]{2014A&A...569A..85S} and constrain nonlocal thermodynamic equilibrium (non-LTE) models \citep{2015ApJ...800L..13H,2019A&A...624A..56V,2019A&A...625A..30L,2021A&A...653A...5P}. Similar studies have yet to be conducted for AR filaments.

At the other end of the wavelength spectrum, low-resolution ($>$20\arcsec) radio observations have shown that quiescent filaments appear as depressions of the background brightness temperatures on disk at continuum wavelengths between 3--8\,mm, possibly due to free-free absorption by dense, cool material, and as emission features at the limb \citep{1972SoPh...25..108K,1979SoPh...61..335R,1981SoPh...71..311S,1992SoPh..137...67V}. At wavelengths around 1\,mm, prominences are still well visible at the limb \citep{1993A&A...274L...9H}, but they are practically invisible on disk \citep{1993ApJ...418..510B,1994IAUS..154...85L}. 
Under certain conditions, brightness measurements at different mm wavelengths can be used to constrain the optical thickness and kinetic temperature of such structures \citep{2015SoPh..290.1981H,2017SoPh..292..130R,2018ApJ...853...21G}. 

The Atacama Large Millimeter/submillimeter Array \citep[ALMA,][]{2009IEEEP..97.1463W} provides an opportunity to investigate filament substructure in the (thermal) millimeter continuum at a much higher spatial resolution than before \citep[see review by][]{2016SSRv..200....1W}. These capabilities have been recently demonstrated for a prominence at the limb \citep{2022arXiv220212761H,2022arXiv220212434L}. The only ALMA observations of filaments available until now have been taken with single dishes with a beam size of approximately 25\arcsec~at 1.25\,mm \citep{2017SoPh..292...88W,2017A&A...605A..78A,2018A&A...613A..17B}. 
Here, we use interferometry maps with a significantly improved spatial resolution ($\sim$\,0.6\arcsec~at 1.25\,mm) that show considerably lower brightness temperatures in an AR filament compared to the background. Because of their smaller spatial scales, AR filaments had not yet been observed in this wavelength range. We compare the ALMA maps to UV imagery provided by IRIS and the Solar Dynamics Observatory \cite[SDO,][]{2012SoPh..275....3P} and magnetograms obtained by the Solar Optical Telescope \citep[SOT,][]{2008SoPh..249..167T} onboard Hinode. 

\section{Observations}

We observed NOAA AR 12738 on April 13, 2019, with ALMA Band 6 (240 GHz or 1.25\,mm) in two execution blocks between 14:15-15:10\,UT and 16:52-17:47\,UT in mosaic mode and with Band\,3 (100\,GHz or 3\,mm) in two execution blocks between 18:19-18:54\,UT and 19:15-19:50\,UT in single-pointing mode. The target was a group of pores and plage region near the disk center at $\mu\approx0.98$ (the cosine of the heliocentric angle), west of a leading sunspot, where an AR filament was also visible for most of the observing campaign. 
We obtained co-temporal observations with IRIS and Hinode for the first Band\,6 block and part of the second block. There are no IRIS nor Hinode supporting observations to the ALMA Band\,3 data.
We also use full-disk images in the ultraviolet range taken by the Atmospheric Imaging Assembly \cite[AIA,][]{2012SoPh..275...17L} and continuum images at 6163\,\AA~obtained by the Helioseismic and Magnetic Imager \cite[HMI,][]{2012SoPh..275..207S} on SDO. We use one \ion{H}{$\alpha$} image taken by the Global Oscillation Network Group \citep[GONG,][]{2011SPD....42.1745H}.

\subsection{Data reduction and calibration}

The ALMA array consisted of 9\,$\times$\,7m and 40\,$\times$\,12m functioning antennas, with a maximum baseline length of about 700\,m. In order to cover a significant region of plage, a mosaic of ten different pointings was carried out using Band\,6, covering a region slightly larger than 1 arcminute across, which is comparable to the Band\,3 field of view (FOV). The FOV of a 12\,m antenna is 25\arcsec~at 230\,GHz, and the centers of adjacent mosaic fields have a separation of 12\arcsec. The observation cycled through the ten mosaic fields in order, acquiring three 2\,s integrations at each visit. The duration of a full mosaic cycle was 85\,s; these were repeated during scans on the Sun approximately 7-min long alternating with 2-min calibration scans.

We attempted to make time-resolved images of complete mosaic cycles using the standard Common Astronomy Software Applications  \citep[\texttt{CASA},][]{2007ASPC..376..127M} package, applying the recommended "mosaic" gridding option, but we found that most of the resulting images were corrupted by bright or dark features at the edges of the mosaic. This did not occur when the full-time range was mapped. To get around this problem, we mapped and self-calibrated every integration in every field separately without primary beam correction, including the four base bands at 230, 232, 245 and 247\,GHz to improve $uv$ coverage, and then combined them in a  linear mosaic in the image plane, weighting overlapping regions of the image using the appropriate primary beam response for the 12\,m antennas. The single-field images were restored with a Gaussian beam with full-width-at-half-maximum of 0.6\arcsec. The three consecutive integrations at each pointing were combined, and at each step, the time-dependent mosaic image was created by replacing the previous image of the current field with the new image, appropriately weighted by the primary beam, and recalculating the mosaic. This resulted in a sequence of 280 mosaic images for each of the two observations with a time separation of 9\,s between mosaics within a scan, but the true time resolution is more complicated due to the overlapping nature of the fields. Comparison of the resulting images with cases where \texttt{CASA} successfully imaged a full 85\,s mosaic cycle indicated good agreement.

The Band 6 mosaics were scaled to $T_{\rm b}$ units, and a value of 6\,600\,K was added to the mosaics to compensate for the background level not observable by the interferometer, derived by inspection of the corresponding FOV in single-dish images (taken with the total-power array) that measure the full solar disk temperatures. The typical temperature range in individual mosaic images is from $\sim$\,4\,000 to $\sim$\,8\,500\,K.

The Band\,3 data consists of single-pointing maps at 2-s cadence, maximum spatial resolution of 1.2\arcsec, and noise level of 20\,K. We refer to \citet{2020A&A...643A..41D} for a detailed explanation of the data reduction, which includes self-calibration, primary beam correction, and absolute flux calibration.

The SDO level 1 data were converted into level 1.5 using the \texttt{aia\_prep} routine in SolarSoftWare \citep[\texttt{SSW},][]{1998SoPh..182..497F}. Further processing including coaligment, resampling, and derotation was performed using the \texttt{SunPy} package \citep{sunpy_community2020}. The HMI 6173\,\AA~continuum images were deconvolved using the \texttt{Enhance} deep learning code\footnote{\url{https://github.com/cdiazbas/enhance}} \citep{2018A&A...614A...5D}. The AIA\,304\,\AA~images were corrected for stray light using semi-empirical point-spread functions \citep{2013ApJ...765..144P} and instrumental degradation. The data cadences are 12\,s and 24\,s in the EUV and UV.

The IRIS data comprises a time series of dense 64-step raster scans in the near-UV and far-UV wavelength ranges. Here we only use the spectral data around the \ion{Mg}{II} h and k lines that probe the chromosphere; there is no signal in the far-UV lines in the filament.
The pixel scale is 0.166\arcsec~along the slit and the size of the FOV is 22\arcsec$\times$67\arcsec. The integration time was 1\,s and the raster cadence was about 2.3\,min. The level 2 data were converted from data units into intensity in physical units using version five of the spectrograph effective areas obtained through the \texttt{iris\_get\_response} routine in \texttt{SSW}.

The Hinode/SOT data consists of level 2 products, which include the magnetic field strength ($\lvert \boldsymbol{B}\rvert$), the stray-light filling factor ($1-f$), inclination ($\gamma$), and azimuth ($\phi$) angles derived from Milne-Eddington inversions of the \ion{Fe}{I} 6301.5 and 6302.5\,\AA~Zeeman sensitive lines \citep{2013SoPh..283..601L}. The line of sight (LOS) component of the magnetic field vector is given by $B_{\rm LOS}=(1-f)\lvert \boldsymbol{B}\rvert\cos(\gamma)$.
The slit step is $\sim$\,0.297\arcsec~and the pixel scale is $\sim$\,0.319\arcsec~along the slit. The size of the FOV is approximately 60\arcsec$\times$61\arcsec. The exposure time was 1.6\,s and the raster cadence was $\sim$13\,min. 

The co-aligment of the IRIS and Hinode data was improved by cross-correlation with the HMI continuum images. The ALMA Band\,6 maps were cross-correlated with AIA 304\,\AA~images taken at an instant close to the middle of each ALMA mosaic sampled at the same spatial resolution. This allowed a proper coalignment of the EUV and mm brightenings recurring near the pore at the center of the FOV \citep[not displayed; see also][]{2020A&A...643A..41D}.

\subsection{Data clustering}
\label{Section:clusters}

To investigate the spatial correlation between the \ion{Mg}{II} h and k profiles and the mm continuum emission, we applied an agglomerative hierarchical clustering method to the former using the \texttt{hierarchy} library in \texttt{Scipy} \citep{2020SciPy-NMeth} wherein profiles are grouped based on a distance metric. We used Ward's linkage method, which minimizes the within-cluster variance \citep{01621459.1963.10500845}, that is the spread in the euclidean distances between each cluster member and the centroid. This technique can be interpreted as a type of dimensionality reduction in that the information in the IRIS spectra with hundreds of wavelength points is condensed into a single number, which can then be compared to the 1.25 mm continuum brightness.
The number of clusters was decided upon visual inspection of the dendrogram. While that choice is arbitrary, we concluded that 15 clusters were a good compromise between limiting the impact of noise and being able to identify sufficient spatial structure within the filament while tracking its time evolution. The clustering is based on 20475 spectral profiles of the wavelength window between 2791.2--2809.6\,\AA, which includes the \ion{Mg}{II} h, k, and UV triplet lines.

\begin{figure}[t]
\begin{center}
\includegraphics[width=\linewidth]{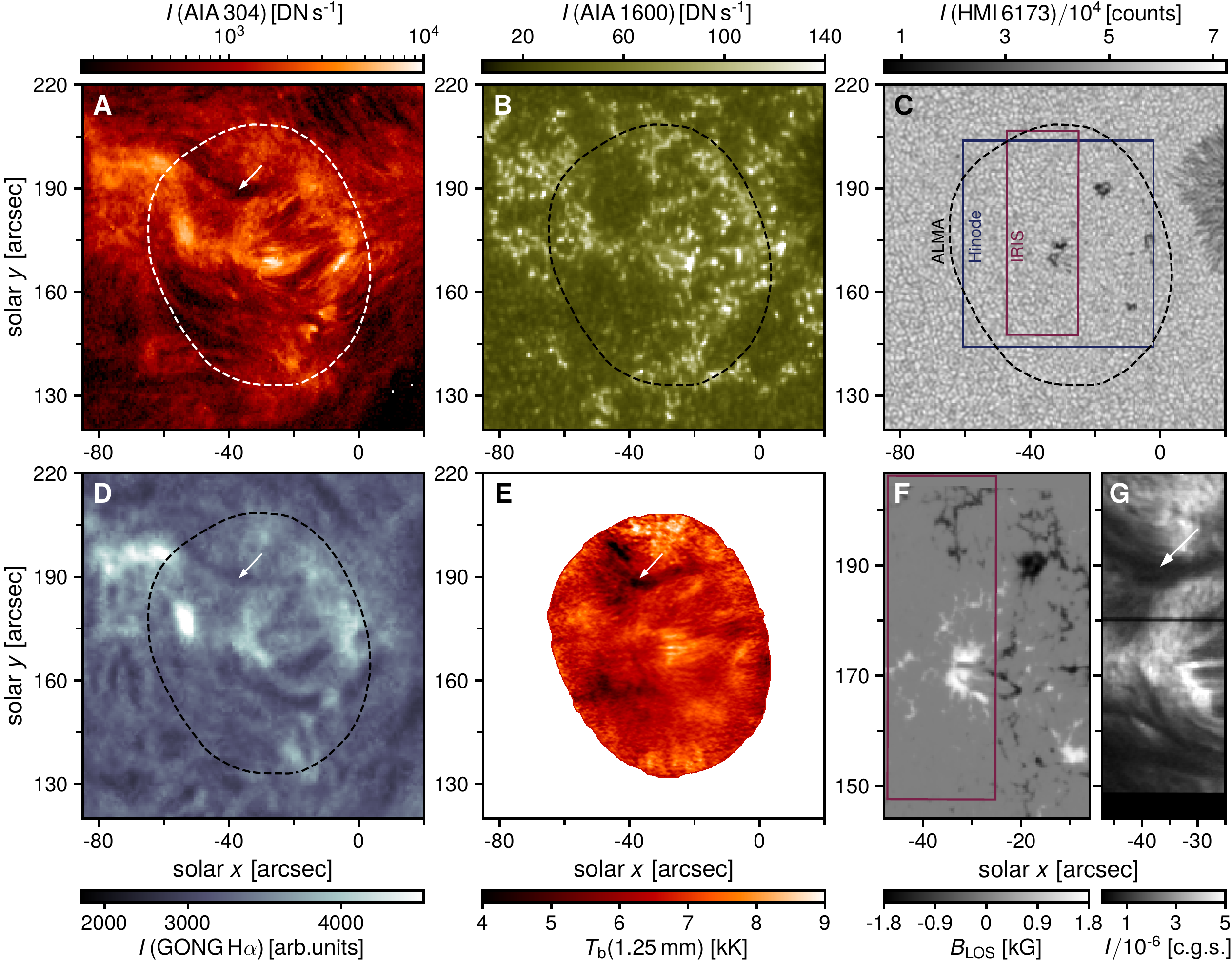}
\end{center}
\caption{Overview of NOAA AR 12738 by different instruments on April 13, 2019, around 14:15 UT. Panel A: SDO/AIA\,304\,\AA~intensity in logarithmic scale; panel B: SDO/AIA\,1600\,\AA~intensity (capped); panel C: deconvolved SDO/HMI 6173\,\AA~continuum intensity; panel D: gamma-adjusted and unsharpened GONG \ion{H}{$\alpha$}; panel E: ALMA Band 6 brightness temperature mosaic; panel F: Hinode/SOT line-of-sight magnetogram; panel G: IRIS intensity in the core of \ion{Mg}{II}\,k (in units of $\rm erg\,s^{\text{-}1} sr^{\text{-}1} cm^{\text{-}2} Hz^{\text{-}1}$). The dashed lines show the ALMA field of view, and the boxes in panel C show the area covered by the Hinode and IRIS raster scans. The arrows indicate the location of an AR filament.} \label{fig:1}
\end{figure}

\section{Results}
\label{Section:OBS_results}

Figure \ref{fig:1} shows the target as observed by the different instruments at the start of the coordinated campaign. Observations of this AR with ALMA have shown that $T_{\rm b}(\rm 3\,mm)$ correlates poorly with AIA\,1600\,\AA~intensity \citep[$r$\,$=$\,0.34, ][]{2020A&A...643A..41D}. Here, we find that $T_{\rm b}(\rm 1.25\,mm)$ correlates slightly better with AIA\,1600\,\AA~emission ($r$\,$=$\,0.45), which is consistent with lower formation heights of shorter mm wavelengths. An overall better agreement is found with the AIA\,304\,\AA~intensities, especially during small-scale transient brightenings similarly to the ones observed with Band\,3 \citep{2020A&A...643A..41D}, but we did not quantify this correlation as the main focus of this paper is the AR filament described below.

The observations also reveal a dark filamentary structure that can be identified in all of the chromospheric diagnostics in the upper part of the FOV. However, the filament appears fragmented into pieces at 1.25\,mm in contrast with the more uniform dark structures seen in \ion{H}{$\alpha$}, \ion{He}{I}\,304\AA, and \ion{Mg}{II}\,k. Brightness values as low as $T_{\rm b}(\rm 1.25\,mm)\sim2800$\,K can be found at certain locations, but other parts of the filament cannot be distinguished from the background, suggesting either significant opacity variations or temperature variations if the filament threads are optically thick. Other dark \ion{H}{$\alpha$} fibrils near the center of the ALMA FOV appear to have no counterpart at 1.25\,mm.

\begin{figure}[t]
\begin{center}
\includegraphics[width=0.98\linewidth]{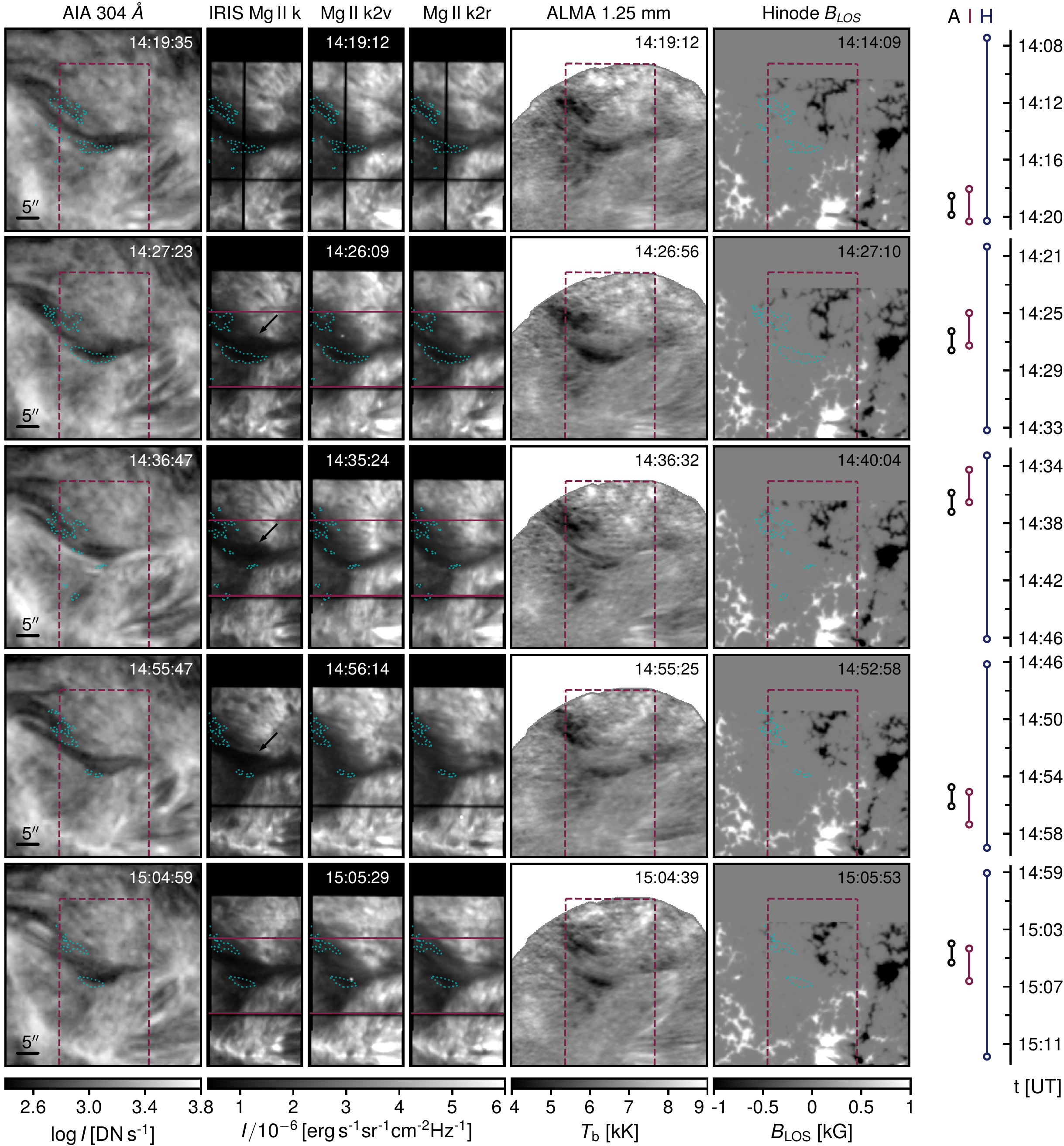}
\end{center}
\caption{Time evolution of the AR filament. The AIA 304\,\AA~images have been averaged within the time span needed by ALMA to mosaic the filament. The intensities are capped for display purposes. The time stamps on the IRIS and Hinode raster images correspond to the central slit position. The dashed boxes show the IRIS FOV. The cyan contours show $T_{\rm b}\rm(1.25\,mm)=5000$\,K. The horizontal magenta lines delimit the area displayed in Fig.\,\ref{fig:3}. The data acquisition times are shown on the right.} \label{fig:2}
\end{figure}

Figure \ref{fig:2} shows the temporal evolution of the filament as seen by the different instruments. The $y$-axes in all panels are aligned with solar north. The appearance of the filament in the Band\,6 maps changes dramatically over one hour as parts of it practically vanish and reappear, whereas the filament spine remains well visible in the \ion{Mg}{II} core images throughout. The dark Band\,6 structures thin out around 14:36\,UT when a bright thread forms in the 304\,\AA~images, following a brightening at the eastern footpoint of the filament.
The lowest $T_{\rm b}(\rm 1.25\,mm)$ values ($<$5000\,K) are located near the widest part of the filament but usually offset from the darkest regions in the 304\,\AA~ images. We find that the coolest $T_{\rm b}(\rm 1.25\,mm)$ values correlate with the darkest features in the k2v and k2r images, whereas the dark clouds in the core of the h and k lines (as indicated by the arrows), which probes higher heights, are indistinguishable from the background at 1.25\,mm. Note that depending on the phase of the IRIS slit-scan, there could be a time lag of up to $\sim$1.5\,min between the latter and ALMA. Even in the darkest part of the filament, the intensities in the core of the \ion{Mg}{II} lines are about 10$-$15\% brighter than the mean QS profile at disk center, yet the filament contrast is high owing to the bright surrounding plage. The SOT magnetograms do not show a significant change in the photospheric magnetic flux underneath the filament spine over time.

\begin{figure}[t]
\begin{center}
\includegraphics[width=\linewidth]{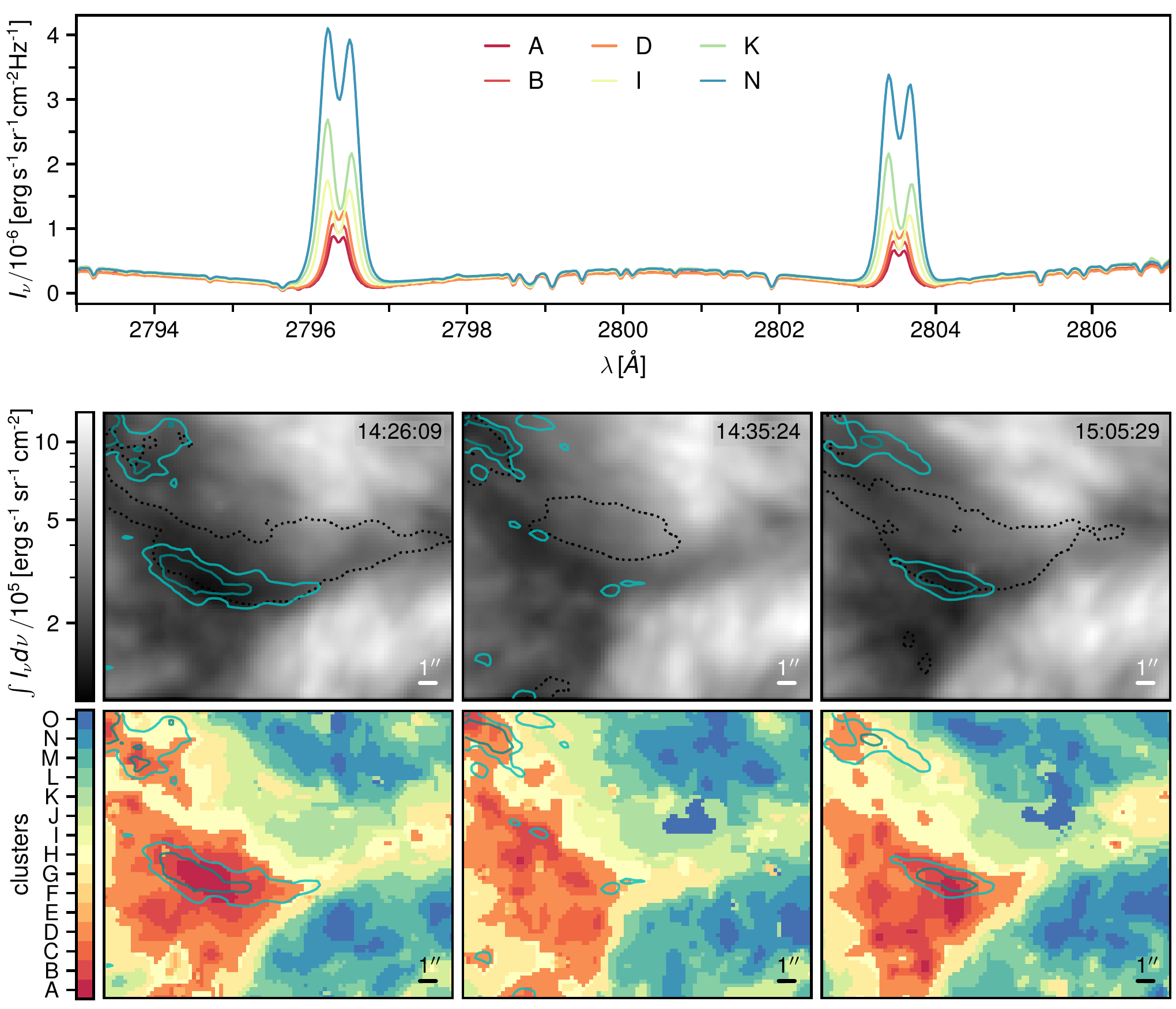}
\end{center}
\caption{Clustering of \ion{Mg}{II} h and k profiles in an AR filament. Upper panel: mean spectral profiles of selected clusters. Middle panel: integrated intensity in the k line in logarithmic scale within the region delimited by the horizontal lines in Fig.\,\ref{fig:2}. Lower panel: clusters of \ion{Mg}{II} h and k profiles. The light and dark blue contours correspond to $T_{\rm b}\rm(1.25\,mm)=5000\,K~\mathrm{and}~4500$\,K, and the dotted black contours outline the filament boundary in the AIA\,304\,\AA~images.
} \label{fig:3}
\end{figure}

Figure \ref{fig:3} shows the results of the clustering algorithm on the IRIS spectra. To create smooth spatial maps of cluster types, profiles belonging to a particular cluster were averaged, and the 15 mean spectra were sorted in terms of integrated intensity (within $\pm 50\rm\,km\,s^{-1}$ from line center) and labeled from A to O; a few of those profiles are displayed for comparison. A simple intensity threshold of 650\,$\rm DN\,s^{-1}$ separates well the filament structure from the background in the AIA\,304\,\AA~passband.
The coolest $T_{\rm b}(\rm 1.25\,mm)$ values ($<$5\,000\,K contours) occur in regions where, on average, the \ion{Mg}{II} h and k lines are the narrowest, the central reversals are the shallowest, and the h and k double peaks are more symmetric, which suggests small velocity gradients. The average k line width measured at half intensity between the averaged k2v and k2r peaks and the averaged k1r and k1v dips is $\approx$\,0.31\,\AA~compared to $\approx$\,0.54\,\AA~in the nearby plage (e.g., profile N). The filament body north of the Band\,6 contours shows \ion{Mg}{II} profiles that are just as dark in the h and k cores as in the lower part (cf. Fig.\,\ref{fig:2}), but they are broader and show stronger, more asymmetrical, and separated k2v and k2r (or h2v, h2r) peaks, so they are classified as a different type. This implies that the coolest Band\,6 temperatures coincide with the regions of lower integrated intensity in the h and k lines where the coolest and densest material may be located. 

The aforementioned trend generally holds at different times but the time lag between the mm and NUV diagnostics has to be taken into consideration. At 14:26, low $T_{\rm b}(\rm 1.25\,mm)$ values coincide with locations with A- and B-type IRIS profiles -- they are essentially of the same kind but B-types are only slightly broader and brighter. These profiles practically disappear at 14:35, so do $T_{\rm b}(\rm 1.25\,mm)<5000$\,K; this appears to be related to the heating event that occurs along the filament spine as shown by the 304\,\AA~image; both signatures reappear at 15:05. 
We note that there are locations outside of the filament that show A- or B-type profiles where low $T_{\rm b}(\rm 1.25\,mm)$ values are not found, and vice-versa, which just indicates relative, spatially-dependent opacity variations in both diagnostics.

\begin{figure}[t]
    \centering
    \includegraphics[width=0.67\linewidth]{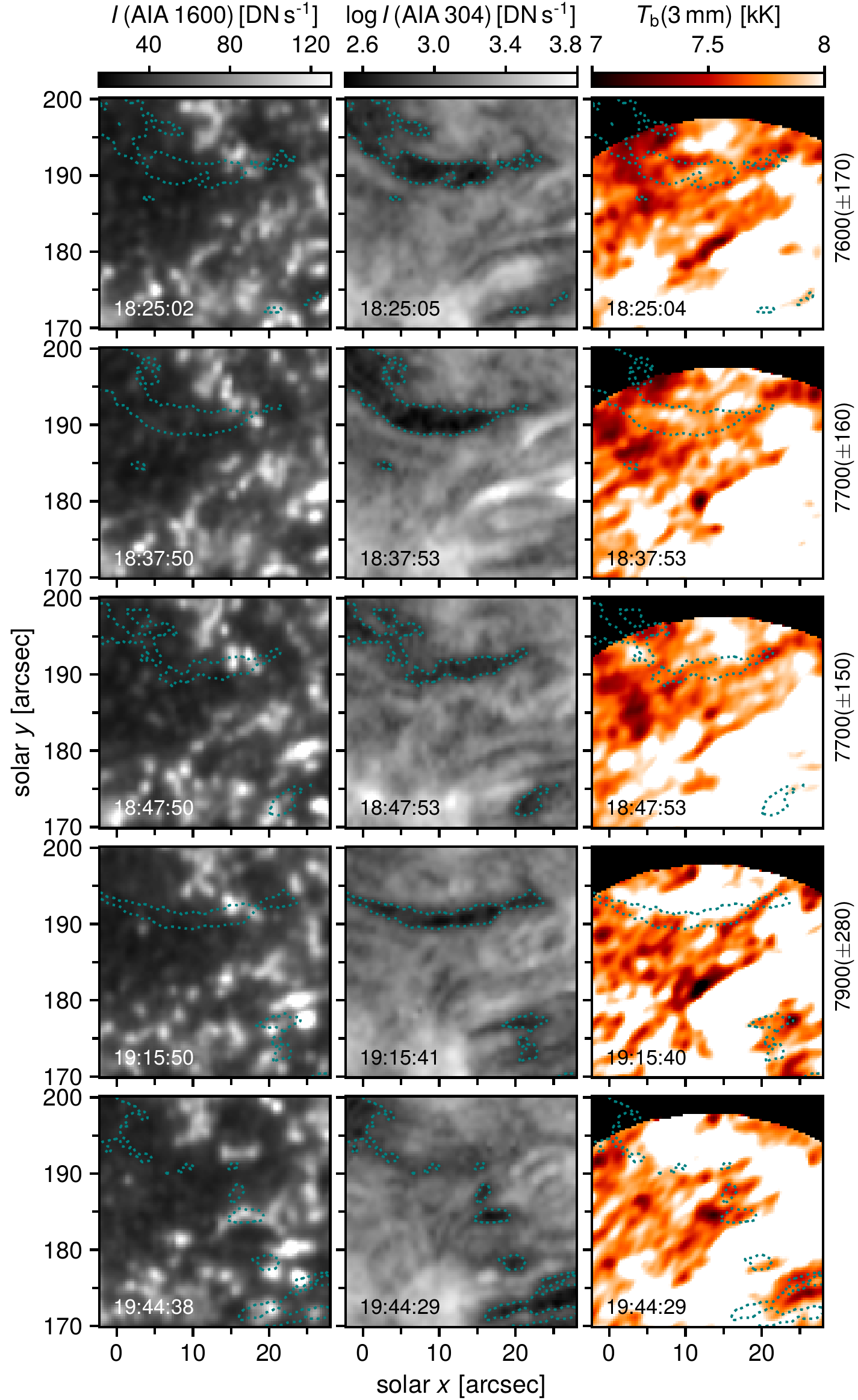}
    \caption{ALMA Band 3 observations at a later time on April 13, 2019. The ALMA colormap range is capped at 8\,kK to enhance small $T_{\rm b}$ variations within the filament. Images in the AIA 1600\,\AA~passband are displayed for context. The intensities are capped for display purposes. The cyan contours delimit the filament structure in the  AIA\,304\,\AA~images. The values on the right indicate the mean($\pm$standard deviation) $T_{\rm b}(\rm 3\,mm)$ within the filament contours.}
    \label{fig:4}
\end{figure}

Figure \ref{fig:4} shows selected ALMA Band\,3 maps obtained after the Band 6 mosaics. 
The AIA images show a more dramatic evolution of the filament compared to an earlier time in the day (cf. Fig.\,\ref{fig:2}), to the point that the filament practically vanished towards the end of the second observing block.
Unlike the 1.25\,mm maps, the 3\,mm maps do not show the same dark features against the background. 
This is interesting in that the optical thickness is expected to increase with wavelength, so the absorbing features in Band\,6 should appear even darker in Band\,3, that is the contrast is expected to slightly increase at longer wavelengths by extrapolation of the contrast curve obtained from previous low-resolution observations of quiescent filaments in the mm and cm ranges \citep[][]{1979SoPh...61..335R}. Therefore, it is somewhat surprising that the filament shows higher contrast in Band\,6 than in Band\,3. However, the dynamic nature of AR filaments may play a role here.
The mean $T_{\rm b}(\rm 3\,mm)$ values within the filament identified in the 304\,\AA~images are around 7700\,K with a few hundred-kelvin variations, which is slightly higher than the average QS level \citep[$\sim$\,7500($\pm100$)\,K,][]{2017SoPh..292...88W}, but these values may be contaminated by the bright plage background because the Band 3 contours overlap with the plage as shown by the AIA\,1600\,\AA~images. We note that the non-simultaneity of the Band\,6 and Band\,3 maps is relevant for interpreting relative brightness ratios.

\clearpage
\section{Discussion and conclusions}

This paper presents ALMA observations of an AR filament at two mm wavelengths supported by SDO and partly by IRIS and Hinode. The filament shows substantial temporal variability in the AIA\,304\,\AA~passband, the \ion{Mg}{II} h and k lines, and in ALMA Band\,6. Brightness variations are also detected in Band\,3 but they are more difficult to discern from the background fluctuations.
As anticipated, the ALMA Band\,6 interferometric $T_{\rm b}$ maps show dark/cool features with much higher contrast relative to the background than the single-dish observations reported thus far \citep{2017A&A...605A..78A,2018A&A...613A..17B}. Very fine dark threads can be seen at certain times (e.g., Fig.\,\ref{fig:2}, 14:36\,UT), but they do not last more than a few minutes. However, not all of the filament body (as seen by IRIS or AIA 304\,\AA) appears dark, but there are significant brightness variations across the filament spine. We note that the filament is also visible in the second Band\,6 observing block until 17:47 UT (not displayed).
We did not identify significant changes in the photospheric magnetic field that we could relate to the filament variability, but we would need field extrapolations to study the topology of the filament itself to investigate a possible correlation between "magnetic dips" and the mm continuum brightness distribution, similarly to what has been done based on \ion{H}{$\alpha$} observations \citep[e.g.,][]{1999A&A...342..867A,2004ApJ...612..519V,2010ApJ...714..343G,2017ApJ...844...70L}.

We find a spatial association between the integrated k line intensity and $T_{\rm b}\rm (1.25\,mm)$ in the filament that is consistent in time. This suggests that both diagnostics are coupled through changes in temperature and electron density. The \ion{Mg}{II} lines are narrower and the central reversals are shallower than the average QS profile at those locations.
The coolest $T_{\rm b}\rm (1.25\,mm)$ values coincide with the dark structures that are visible away from the k (and h) line center and thus extend down to lower heights in the atmosphere. 
The lower atmosphere may be filled with low-lying dense cool threads where the absorption is high but the emissivity is low, hence the weak \ion{Mg}{II} lines. However, interpreting the \ion{Mg}{II} line shapes is not trivial as they depend on the interplay between the incident radiation, gas pressure, filament thickness, and the properties of the prominence-corona transition region \citep{2014A&A...564A.132H}, and thus require a follow-up investigation. 

Detailed radiative transfer models of AR filaments are scarce in the literature. The spectral synthesis based on a 3D whole-prominence fine structure model presented by \citet{2016ApJ...833..141G} shows filament threads in emission by a few hundred kelvins above a uniform disk background at 1.25\,mm unlike what we observed; the simulated filament is optically thin ($\tau_{\nu}<1$) at this wavelength.
The emerging brightness temperature is the result of the absorbed background brightness temperature, $T^{\rm bcg}_{\rm b}$, plus the integrated emission along a geometrical path of length $L$:
\begin{equation}
    T_{\rm b}(\nu) = T^{\rm bcg}_{\rm b}(\nu)\,e^{-\tau_{\nu}}+\int^{L}_{0}T\kappa_{\nu}\,e^{-\int^{l}_{0}\kappa_{\nu}dl^{\prime}}\,dl, \label{eq:1}
\end{equation}
\noindent where $\kappa_{\nu}$ is the frequency-dependent absorption coefficient \citep[e.g.,][]{2016ApJ...833..141G}. Brightness temperature values lower than the background level require kinetic temperatures, $T$, lower than background temperatures. However, the observed AR filament and background certainly have different properties than the simulated ones. Dedicated simulations with more realistic AR conditions are needed to investigate this.

Constraining both the kinetic temperature and thickness of the prominence is not possible using single-band data but it requires simultaneous observations at two wavelengths and the assumption of isothermal conditions \citep[e.g.,][]{2015SoPh..290.1981H,2017SoPh..292..130R}. In that case, Eq.\,\ref{eq:1} simplifies to
\begin{equation}
    T_{\rm b}(\nu) = T^{\rm bcg}_{\rm b}(\nu)\,e^{-\tau_{\nu}}+T(1-e^{-\tau_{\nu}}).
\end{equation}
An hour after the Band\,6 observations, the filament is still visible in the AIA\,304\,\AA~images, but it is indiscernible from the background in Band\,3. This cannot be attributed to a difference in spatial resolution (by a factor of $\sim$2) because the dark threads in the Band\,6 with the typical sizes of a few arcseconds would be well resolved with Band\,3. Instead, it could imply that the filament temperature is similar to the background. We note that the background is far from uniform, which makes it complicated to interpret $T_{\rm b}$ variations within the filament.
Assuming that the average properties of the filament did not significantly change between the time the Band\,6 and Band\,3 maps were taken, which is a reasonable assumption at least for the first Band\,3 observing block (Fig.\,\ref{fig:4}), it is difficult to reconcile these observations with an isothermal model since we would expect even stronger absorption features at 3\,mm than at 1.25\,mm \citep[see also][]{2019ApJ...875..163R}. Following \citet{1985ARA&A..23..169D}, the opacity ratio in both bands is given by
\begin{equation}
    \frac{\kappa_{3}}{\kappa_{1.25}} = \left[\frac{\nu_{1.25}}{\nu_{3}}\right]^{2}\frac{18.2+\ln(T^{3/2})-\ln \nu_{3} }{18.2+\ln(T^{3/2})-\ln \nu_{1.25} },~ (T<2\times10^{5}\,\mathrm{K}).
\end{equation}
An optically thick plasma at 3\,mm with a kinetic temperature of 7500\,K (equal to the mean QS background at 3\,mm) would show a higher mean $T_{\rm b}(\rm 1.25\,mm)$ than what we have observed. For example, $\tau_{\rm 3\,mm}=1$ implies $\tau_{\rm 1.25\,mm}=0.15$ and $T_{\rm b}(\rm 1.25\,mm)$\,$\sim$\,6100\,K, which would make the filament stand out above the background. Therefore, these two wavelengths are more likely sensitive to different layers of the filament with different temperature and density conditions, or there is a temperature gradient spanning their formation heights, with Band\,6 probing closer to the lowest temperatures of the filament, and Band\,3 sensing warmer filament plasma or possibly the outermost shell that separates it from the corona if the optical thickness is large. Simultaneous observations of AR filaments in both ALMA bands are needed to clarify these findings.

Combined non-LTE radiative transfer modeling of the \ion{Mg}{II} h and k lines and the radio continua will be required to diagnose the properties of the observed filament. Using such an approach, ALMA observations will allow us to constrain the plasma temperature, while the IRIS spectra can provide information about the pressure/density and the LOS velocity gradients.

\bibliographystyle{frontiersinSCNS_ENG_HUMS} 

\section*{Conflict of Interest Statement}
The authors declare that the research was conducted in the absence of any commercial or financial relationships that could be construed as a potential conflict of interest.

\section*{Author Contributions}

JS, SW, and JL contributed to the ALMA research proposal and data acquisition. SW performed the ALMA data reduction. JS conducted the analysis. JS, SW, KR, GC, SG, PH, and JL contributed to the discussion and writing.

\section*{Funding}

SG and PH acknowledge support from the grants 19-16890S and 19-17102S of the Czech Science Foundation (GAČR). SG and PH thank for the support from project RVO:67985815 of the Astronomical Institute of the Czech Academy of Sciences.

\section*{Acknowledgments}
This paper makes use of the following ALMA data: ADS/JAO.ALMA\#2018.1.01518.S. ALMA is a partnership of ESO (representing its member states), NSF (USA) and NINS (Japan), together with NRC (Canada), MOST and ASIAA (Taiwan), and KASI (Republic of Korea), in cooperation with the Republic of Chile. The Joint ALMA Observatory is operated by ESO, AUI/NRAO and NAOJ. IRIS is a NASA small explorer mission developed and operated by LMSAL with mission operations executed at NASA Ames Research center and major contributions to downlink communications funded by ESA and the Norwegian Space Centre. Hinode is a Japanese mission developed and launched by ISAS/JAXA, with NAOJ as domestic partner and NASA and STFC (UK) as international partners. It is operated by these agencies in co-operation with ESA and NSC (Norway). Data were acquired by GONG instruments operated by NISP/NSO/AURA/NSF with contribution from NOAA. The NSO is operated by the Association of Universities for Research in Astronomy, Inc., under cooperative agreement with the National Science Foundation. This research has made use of \texttt{Astropy} (\url{https://astropy.org}) -- a community-developed core Python package for Astronomy \citep{astropy:2018}, \texttt{SunPy} (\url{https://sunpy.org}) -- an open-source and free community-developed solar data analysis Python package \citep{sunpy_community2020}, and MATLAB and the Image Processing Toolbox release 2021b (The Mathworks, Inc., Natick, MA, USA).

\section*{Data Availability Statement}

The SDO data can be obtained from the Joint Science Operations Center (\url{http://jsoc.stanford.edu}).
The IRIS data can be downloaded from the Heliophysics Events Knowledgebase Coverage Registry (\url{https://www.lmsal.com/hek/hcr}).
The Hinode data can be obtained from the Community Spectro-polarimetric Analysis Center (\url{http://www2.hao.ucar.edu/csac}). The GONG data can be fetched from Data Archive (\url{https://gong2.nso.edu/archive}).
The raw ALMA data can be found at the ALMA Science Archive (\url{https://almascience.nrao.edu/aq}).

\end{document}